\documentclass[applsci,article,accept,pdftex,moreauthors]{Definitions/mdpi}

\firstpage{1}
\pubvolume{13}
\issuenum{5}
\articlenumber{3242}
\pubyear{2023}
\copyrightyear{2023}
\externaleditor{Academic Editor: Yu-Dong Zhang}
\datereceived{26 January 2023}
\daterevised{27 February 2023} 
\dateaccepted{27 February 2023}
\datepublished{3 March 2023}
\hreflink{https://doi.org/10.3390/\linebreak app13053242} 

\Title{A Novel Implicit Neural Representation for Volume Data}

\TitleCitation{A Novel Implicit Neural Representation for Volume Data}

\Author{{Armin Sheibanifard} 
*$^{,\dagger}$\href{https://orcid.org/0000-0003-2317-0267}{\orcidicon} and Hongchuan Yu $^{\dagger}$}

\AuthorNames{Armin Sheibanifard and Hongchuan Yu}

\AuthorCitation{Sheibanifard, A.; Yu, H.}
\address[1]{{National Center for Computer Animation, Faculty of Media and Communication, Bournemouth University, Bournemouth~BH12~5BB,~UK}}
\corres{\hangafter=1 \hangindent=1.05em \hspace{-0.82em}Correspondence: asheibanifard@bournemouth.ac.uk}

\firstnote{\hangafter=1 \hangindent=1.05em \hspace{-0.82em}These authors contributed equally to this work.}

\abstract{The  storage of medical images is one of the challenges in the medical imaging field. There are variable works that use implicit neural representation (INR) to compress volumetric
medical images. However, there is room to improve the compression rate for volumetric medical images. Most of the INR techniques need a huge amount of GPU
memory and a long training time for high-quality medical volume rendering. In this paper, we
present a novel implicit neural representation to compress volume data using our proposed architecture, that is, the Lanczos downsampling scheme, SIREN deep network, and SRDenseNet high-resolution scheme. Our architecture can effectively reduce training time, and gain a high compression rate while retaining the final rendering quality. Moreover, it can save GPU memory in comparison with the existing works. The experiments show that the quality of reconstructed images and training speed using our architecture is higher than current works which use the SIREN only. Besides, the GPU memory cost is evidently decreased.}


\begin{document}

\section{Introduction}

Contemporary microscopy technology is widely applied to biology, neuroscience,  and medical imaging fields, which bring about large-scale multidimensional datasets and require terabytes or petabytes of data storage. Besides, recent developments in radiological hardware have increased the capabilities of medical imaging to produce high-resolution and 3D images that require high storage capacity. It is desired to manipulate multidimensional datasets for visualization and predictions from multi-channel image intensity values in a 2D or 3D scan. Such datasets pose significant challenges related to storage, manipulation,  and rendering. In this paper, we suggest a novel implicit neural representation to compress high-resolution medical volume data using the implicit neural representation (INR) network with high speed and quality in comparison with current works.

There are some works using INR with periodic activation functions (SIREN)~\cite{SIREN} to compress the medical images~\cite{COM1,COM2}. In order to compress the images, these techniques train a neural network using the voxel coordinates. The resulting trained networks represent the individual image stacks. A distinct advantage is that the size of the trained deep network is always less than that of the image stack. However, they also suffer some considerable issues. They usually depend on a huge GPU memory, specifically for high-resolution volume data. Thus, it is impossible for clinicians to access medical volume data based on usual workstations and laptops. On the other hand, in order to deal with high-resolution (HR) image stacks, it is natural to use large-scale neural networks with multiple layers, which results in a long training time and a low rate of compression. Obviously, the main challenge is to both keep a high rate of compression to decrease deep network size and keep the quality of reconstructed images.

To tackle this challenge, we present an architecture combining the SIREN to compress the volume data effectively and a super-resolution module to keep the quality of the reconstruction. Our architecture includes a downsampling module to decrease the resolution of high-resolution (HR) and low-resolution (LR) images, which can fit the downsampled volume data in usual GPUs with low graphical memory. As a result, the  SIREN's layers can be reduced. Training can be accelerated accordingly. The following super-resolution (SR) reconstruction network can recover images up to the quality of the original HR.

In summary, our contributions include:
\begin{itemize}
\item Our architecture consists of the Lanczos downsampling scheme, SIREN deep network, and SRDenseNet upsampling scheme, which increase the speed of training and decrease the demand for GPU memory in comparison with existing INR-based compression techniques;
\item Our architecture can reach both a high compression rate and high quality of the final volume data rendering.
\end{itemize}

The following part of this paper is organized as follows. Section \ref{222} presents the related work. Section \ref{333} illustrates the methodology. Section \ref{444} shows the experiments and results from the analysis. Section \ref{555} draws our conclusions and presents future works.

\section{Related Work}\label{222}
Recently, implicit neural representation (INR) is considered by variable works to represent medical imaging. Three-dimensional medical imaging is usually considered as a discrete grid of voxels. Although voxel-based techniques are simple and regular to use, they are used for small voxel grids. Some works try to overcome this limitation and increase the size of voxel grids, using shallow networks with a small batch size~\cite{Voxel} which increases the time of the~training.

\subsection{Implicit Neural Representation}
As the memory consumption of discrete voxel grids increases cubically, the Ref.~\cite{IOSNet} suggests an implicit organ segmentation network, using continuous implicit neural representations. As the IOSNET is a continuous function, it is completely independent of spatial resolution. Because of its continuity, high-resolution medical images can be processed quickly using IOSNET.
The Ref.~\cite{DeepSDF} suggests DeepSDF as a fully continuous and implicit representation for generative 3D modelling. DeepSDF is based on Sign Distance Function, but it uses a generative model to produce 3D continuous interfaces. DeepSDF reduces memory consumption in comparison with its counterparts. The suggested auto-encoder in DeepSDF needs explicit optimization which consumes more time during inference.
In the Ref.~\cite{TSDF} they introduced an encoder–decoder neural architecture to losslessly compress truncated signs in distance fields (TSDF) in a 3D voxel grid. Their deep network architecture is block-based that is trained end-to-end. The main limitation of their model architecture is that the blocks are independent and identically distributed. Ignoring this limitation may help to increase the rate of compression. A lossless compression technique (MedZip) using Long Short-Term Memory (LSTM) was introduced by the Ref.~\cite{LSTM}. Their work predicts the next intensity value using LSTM in a set of the voxels' neighbourhood concept. MedZip is the first lossless compression technique that uses LSTM for volumetric MRI and CT.

In the Ref.~\cite{NeRF} the authors suggested a technique to synthesize new views of a volumetric scene using implicit neural representation as a continuous function. Their technique represents a volume using a deep fully connected network with five inputs and four outputs. The network encodes the spatial location and direction to RGB and opacity. NeRF is time-consuming and its capability in complex images is not considerable.
Following this work, the Ref.~\cite{SIREN} demonstrates that using periodic activation functions outperforms ReLU-MLPs. They also suggest sinusoidal representation networks (SIRENs) that use periodic activation functions in implicit neural representation. This technique fits complex signals and natural images.
In the Ref.~\cite{Occupancy} a 3D representation technique is presented to reduce the memory footprint. In their design, they use a neural network to predict the occupancy function to obtain a continuous volume.
The Ref.~\cite{COIN} presents a compression technique using an implicit neural network (COIN) to compress natural images. It uses multi-layer perceptron (MLP) to encode geometric inputs. COIN's results are weaker than state-of-the-art compression techniques.
The Ref.~\cite{adversarial} suggests an adversarial generation of continuous images using INRs. Their model is constituted of two techniques, multi-scale INRs and factorized multiplicative modulation (INR-GAN). With this architecture, it is possible to represent high-resolution images. Because high sensitivity of INR-based techniques to high-frequency features, it leads to cause artefacts.
The Ref.~\cite{NeRP} introduces an implicit neural representation technique with prior embedding to reconstruct sampled medical images
~\cite{NeuralScene} suggests a generative network based on implicit neural network representation to reconstruct new viewpoints by learning from different viewpoints from input images taken. The Ref.~\cite{Fourier} suggests using Fourier features in MLP. As a standard MLP is slow in convergence, they suggest passing the inputs of the network through a simple Fourier feature mapping. By this technique, the network can learn high-frequency functions and improve the MLPs' performance.

\subsection{Deep Neural Network in Medical Image Restoration}
In the Ref.~\cite{Deeptransfer} the authors suggest a technique to use transfer learning to reconstruct the accelerated MR images. This work shows the capability of transfer learning for sparse training data.
To accelerate dynamic MRI, the authors in the Ref.~\cite{CRNN-MRI} used a convolutional recurrent neural network (CRNN-MRI). Through this method they reconstructed high-resolution MR image sequences.
To accelerate MR image acquisition, the Ref.~\cite{SAT} introduced a self-attention CNN architecture to reconstruct MRI.
Image quality can be easily degraded by noises and artefacts in low-dose computed tomography (LDCT). To address this issue, the Ref.~\cite{LDCT} proposes a deep iterative reconstruction estimation (DIRE) with a 3D residual convolutional network (ResNet) architecture to improve the quality of images. Researchers in the Ref.~\cite{SART} proposed a simultaneous algebraic reconstruction technique (SART) to reconstruct images for translational CT. Then they used a pre-trained CNN to remove artefacts and noise. To reduce the radiation dose, the Ref.~\cite{Dear} suggests a deep encoder–decoder adversarial reconstruction (DEAR) network to reconstruct 3D CT images directly from real clinical cone beam image data. To extract 3D details from generated slices with an adversarial network, they use DEAR-3D which is based on 3D convolutional layers.
In low-dose CT imaging, noise and artefacts are inevitable. To address that issue, the Ref.~\cite{Googlenet} proposes an improved version of GoogLeNet to remove artefacts that are caused by missing projection during image reconstruction.
To reconstruct an X-ray super-resolution image, the Ref.~\cite{SR} suggests a GANs-based approach. It proposes  spectral normalization super-resolution medical images (SNSR-GAN) to reconstruct high-resolution X-ray images.
To reconstruct an ultra-fast CT image, the Ref.~\cite{MRDC} suggests a multi-receptive field densely connected CNN (MRDC-CNN). They also use dense skip modules instead of simple skip modules to flow the information between the encoder and decoder. To decrease the memory footprint they ignore batch normalization.
The Ref.~\cite{PACT} presents a deep regularization method to overcome the reconstruction of photo-acoustic computed tomography (PACT) with sparse view measurements.
A non-local deep image before reconstructing the Positron Emission Tomography (PET) images is proposed by the Ref.~\cite{DIP}. They use a prior image of the patients as the input of the network. They suggest the 3D U-Net~\cite{3DUNet} as the backbone of their network structure.
DeepPET is introduced by the Ref.~\cite{DeepPET} to reconstruct PET images from sinograms. They use a deep encoder–decoder network to reconstruct high-quality images. DeepPET is 108 times faster than its counterparts.
DUG-RECON~\cite{DUG-RECON} is an Unet-based deep learning pipeline to reconstruct PET and CT images directly. It uses a convolutional generative network with three stages, denoising, image reconstruction and super-resolution segments. To reconstruct dynamic PET images, the Ref.~\cite{DYPET} suggests non-negative matrix factorization (NMF) with a deep image prior. They also show the capabilities of DIP for PET image reconstruction. The Ref.~\cite{DREAMNET} suggests a Deep Residual Error Iterative Minimization Network to reconstruct sparse-view CT. They optimize a hand-crafted function to reconstruct high-quality images. In CNN, the size of the convolutional kernel is smaller than the image size. In this case, they cannot understand the whole of the image. The Ref.~\cite{SSTrans-3D} suggests using vision transformers to remove restrictions of CNNs in image reconstruction. The slice-by-slice transformer network (SSTrans-3D) is a transformer-based technique that reconstructs 3D single-photon emission computed tomography (SPECT) images. In image reconstruction, incomplete projection data can cause considerable artefacts. Researchers in the Ref.~\cite{DIOR} suggest Deep Iterative Optimization-based Residual-learning (DIOR) to reconstruct limited-angle CT. To improve generalization ability, DIOR combines deep learning and iterative optimization.
\subsection{Deep Learning and Super-Resolution (SR) Techniques}
Generally, super-resolution (SR) techniques can be clustered into traditional and deep learning methods. In SRCNN~\cite{SRCNN39}, an SR model is trained based on CNN to reconstruct HR from a given LR image. However, in terms of image quality, there are some limitations in SRCNN models. A huge amount of stacked layers causes gradient-vanishing issues. To solve those problems, researchers in the Ref.~\cite{verydeep40} suggested using a very deep convolutional network base on VGG-net. For some of the deep learning techniques in SR, overfitting is highly likely and models can be so big to be stored. To address these issues, a deeply recursive convolutional network (DRCN)~\cite{DRCN41} is suggested. In DRCN, a convolutional layer is repeated many times and the number of parameters depends on the number of applied recursions. As DRCN uses stochastic gradient descent widely, it cannot converge easily. To overcome the difficulties of the training, they use recursive supervision and skip connection. In the Ref.~\cite{Remote42}, authors suggest an end-to-end deep neural network that is constituted of an encoder, a fusion module and a decoder. The encoder takes the features of the LR image. Then they apply a Gated Recurrent Unit (GRU)-based module to combine the features. The super-resolution image is reconstructed by the decoder. The Ref.~\cite{Remote43} suggests coupled-discriminate GAN (CDGAN). In CDGAN, HR and SR images are taken by a discriminator. The network can learn to discriminate low-frequency images. An unsupervised Image Super-Resolution is suggested in the Ref.~\cite{unsup44}. They propose a framework that is based on two parts, unsupervised translation from the original LR image to the reconstructed LR image and supervised SR between reconstructed LR and HR images. In the Ref.~\cite{MZSR} the authors present Meta-Transfer Learning for Zero-Shot Super-Resolution (MZSR). As CNNs are mostly limited to the trained supervised data, they are applicable to specific images. On the other hand, with CNNs it is not applicable to extract internal details of the images. To solve the problems they suggest using zero-shot super-resolution to train internal information. To overcome several gradient updates, they apply meta-learning for zero-shot super-resolution. In remote sensing, There are some issues in super-resolution reconstruction using deep learning, such as model training difficulties and blurred image edges. The Ref.~\cite{CA} proposes a technique with the combination of residual channel attention (CA) to extract deep features. They combine shallow and deep futures, using skip connections to improve the model training. After training, the super-resolution images have sharper edges. In the Ref.~\cite{RBPNet} authors suggest using a Residual Back-Projection network (RBPNet) to reconstruct SR from extremely low-resolution face images. To generate the low-resolution feature map, RBPNet projects the high-resolution feature map to the low-resolution feature space. Then to make a residual feature map for the low-resolution feature map, they subtract the low-resolution feature map from the original feature map. Finally, to generate high-resolution feature space, the low-resolution feature map is projected into high-resolution feature space. RBPNet produces more precise high-resolution images using residual learning. During the application of SR techniques in low-resolution face images, face structure details cannot be recovered well. Thus, a novel SPatial Attention Residual Network (SPARNet) is suggested by the Ref.~\cite{SPARNET}. SPARNET uses a spatial attention mechanism to enable convolutional layers to focus on key face structures and pay less attention to less important details. The results show that their technique can detect key face structures well for extremely low-resolution ({16} 
{$\times$} 16) face images. They also extend SPARNet to SPARNetHD which can generate super-resolution face images (512~{$\times$}~512). In the Ref.~\cite{FGRDN}, the authors address the heavy computation and the high number of parameters of current SR techniques. They suggest the feedback ghost residual dense network (FGRDN). Instead of the residual dense blocks (RDB) they use ghost modules (GM) which can prevent the rise of the parameters by increasing the network depth. FGRDN can converge faster than other corresponding algorithms. The Ref.~\cite{TE-SAGAN} suggests a texture enhancement and generative adversarial network (TE-SAGAN), to generate super-resolution remote sensing images. They suggest an improved generator which is based on a residual network with self-attention and weight normalization. By improving training model stability, the generator can generate images with higher quality. The Ref.~\cite{CA-FRN} proposes a fused recurrent network via channel attention (CA-FRN). Their main concerns are addressed regarding overfitting and the number of parameters. In SR techniques, by increasing the number of layers, the risk of overfitting increases too. CA-FRN uses a recursive channel attention block to pay attention to high-frequency information.  In their model, high- and low-resolution information is fused to generate better results.

\subsection{Volume Data Compression}
Dealing with volumetric datasets may cause slow representation and huge file sizes. The Ref.~\cite{compression1} suggests that a quadtree encoding-based model compresses the volumetric medical images. Their technique is constituted of three stages, initialization, processing, and variable length encoding. In  comparison to the octree technique, it shows better results in the image compression rate. In the Ref.~\cite{compression2}, authors suggest a 3D hierarchical listless block (3D-HLCK). They utilize a 3D listless technique to compress the volumetric medical images. Their results illustrate that 3D-HLCK outperforms the 3D-SPIHT method. Researchers in the Ref.~\cite{compression3} present a GPU-based compression technique for volumetric medical images. In their work, they propose a caching strategy to render high-resolution volumetric medical images. In the Ref.~\cite{compression4} authors suggest a compression technique based on the wavelet transform domain. Their results show that the reconstruction quality of the volumetric medical images is not considered high. The Ref.~\cite{compression5} presents a volume data compression technique using the regression function. Their technique is based on a multi-layer perceptron (MLP) neural network to compress the volumetric medical images. In the Ref.~\cite{shen} they used a stacked autoencoder to compress the malaria-infected blood cells. Authors in the Ref.~\cite{COM1} used neural networks with periodic activation functions (SIREN) to compress multidimensional medical images. Their results show that SIREN outperforms other INR-based compression techniques with ReLU or tanh activation functions. The Ref.~\cite{Autoencoder} suggests CNN-based image compression to compress malaria cell images using a compressor–decompressor framework. According to their technique, there are two autoencoders, where one learns low-frequency components and another learns high-frequency components. As they use Huffman coding and decoding, it is necessary to have a large-sized training dataset to obtain better results.
\section{Methodology}\label{333}
In the following section, we present our architecture and explain its stages in detail.
\subsection{Our Architecture}
\begin{figure}[H]
\begin{adjustwidth}{-\extralength}{0cm}
\

\centering
\includegraphics[width=16cm]{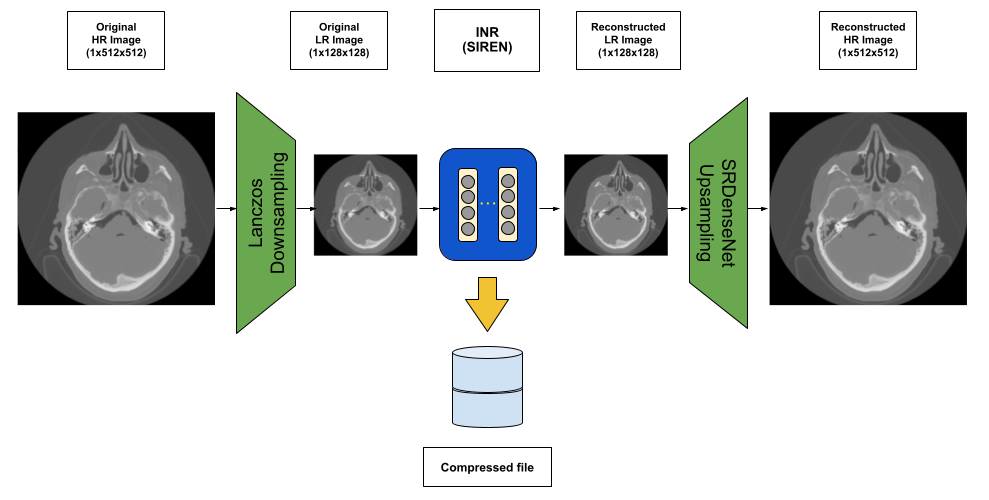}
\end{adjustwidth}
\caption{Our suggested architecture using INR (in this work,  SIREN) to compress high-resolution (HR) medical images.}
\label{fig:Architecture}
\end{figure}

We aim at the medical volume data compression technique to work on usual workstations and laptops instead of the overdependency on high-end GPUs. In order to increase the compression rate and the quality of the reconstruction, we present an architecture that can be applied in many scenarios, as shown in Figure \ref{fig:Architecture}. Our architecture consists of three modules, where the first module can downsample the original high-resolution images using Lanczos to decrease the volume data size. The second module can work on the LR volume data for the implicit neural representation. The third module can take the LR volume data back to the original size using SRDenseNet. As a result, the original volume data can be represented by the neural network to achieve compression purposes. The first module may result in information loss. The third module can recover it. There is no guarantee that it is lossless. However, the experiments show our architecture can reach high PSNR.

\subsection{Lanczos  Resampling}
The Lanczos resampling technique~\cite{Lanczus} is an interpolation technique which is used to resample (downsample/upsample) the volumetric medical images. Here, we use Lanczos resampling to downsample our HR images. In the Lanczos resampling technique, each selected sample is replaced with a translated copy of the Lanczos kernel, which is a limited Sinc function. Equation (\ref{eq:1}) shows the Lanczos kernel formula.
\begin{equation}
L(x) =
\begin{cases}\label{eq:1}
sinc( \pi x)sinc( \pi x/a) & if |x|<a \\ 0 & otherwise.
\end{cases}
\end{equation}

\subsection{Sinusoidal Representation Networks (SIREN)}
We apply the SIREN deep network to medical image stacks. It may be regarded as a function that accepts the 3D coordinate of a {voxel} 
$\mathbf{x} = (x,y,z)$ and outputs the corresponding intensity value $\sigma$. In this case, the neural network becomes a continuous implicit {function} $F_\Theta: \mathbf{x} \rightarrow \sigma$ and can represent a continuous scene of medical image stacks. The training is to optimize its weights $\Theta$ for accurate representation. Recently, INRs have been one of the considerable areas of research focus. Most of the methods adopt a MLP structure using ReLU~\cite{NeRF} as the activation function. ReLU-based MLPs are suffering from the lack of capacity to represent fine details. ReLU networks are linear and their second derivative is zero. As a result, it is impossible to model high-order signals. In the various versions of SIREN, the Sine function is used as the activation function to model fine details~\cite{SIREN}. We applied a small version to training, and the resulting deep network was so small. As a new representation of the original volume data, it effectively compresses the volume data and saves training time.
\subsection{SRDenseNet}
In order to obtain the original resolution image, we use SRDenseNet~\cite{SRDensenet} as a super-resolution module, which is a deep network. We selected SRDenseNet because of its generalization. We note that it can work well on a large class of medical images. For example, we use a small training image set from a large class and then perform it on unseen images of the same class. There are three types of SRDenseNets, SRDenseNet-H, SRDenseNet-HL, and  SRDenseNet-All. SRDenseNet-H uses high-level features to reconstruct HR images. SRDenseNet-HL uses a combination of low-level and high-level features to reconstruct HR images. SRDenseNet-All, which is used in our implementation, uses a combination of all levels of features with densely skipped connections to reconstruct HR images. SRDenseNet-All is constituted of four stages, low-level features, high-level features, deconvolution layers, and a reconstruction layer.

\begin{figure}[H]
\begin{adjustwidth}{-\extralength}{0cm}
\centering
\includegraphics[width=16.7cm]{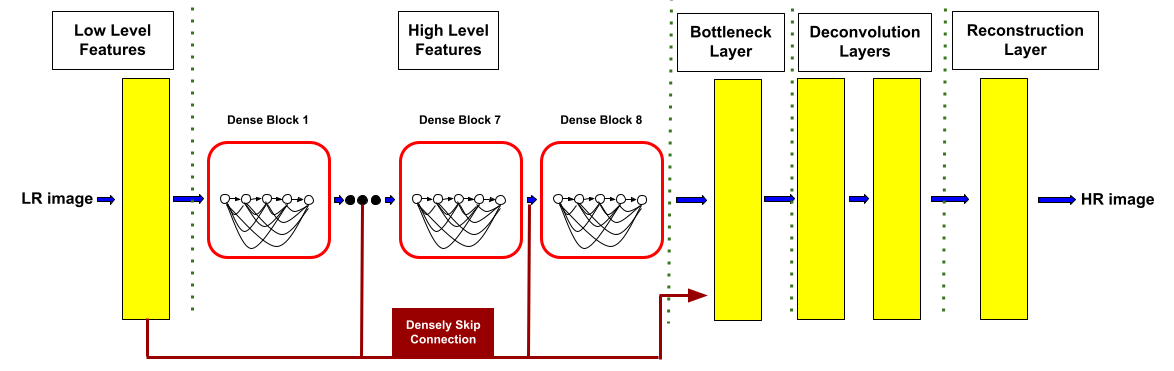}
\end{adjustwidth}
\caption{SRDensenet—all architecture with eight blocks.}
\label{fig:Arch1}
\end{figure}

In the low-level feature stage, the network receives a low-resolution (LR) image as input. A convolution layer learns low-level features. There are eight dense blocks that learn high-level features. Each dense block is constituted of eight convolution layers and each layer produces 16 feature maps. Thus, each block produces 128 feature maps. In the bottleneck layer, the number of feature maps is decreased to a compact model. Then, image resolution is up-scaled from 128~{$\times$}~128 to 512~{$\times$}~512 using a deconvolution layer. Finally, with another convolutional layer with a 3~{$\times$}~3 kernel, the output channels are reduced to a single channel. Figure \ref{fig:Arch1} shows the SRDenseNet-All structure which is used in this paper. In this work, we use the pre-trained model of SRDenseNet in
~\cite{SRDensenet}. We trained the SRDenseNet network with LR and HR pairs of slices of our volumetric medical images dataset. Then a pre-trained model of our training is used in our architecture. According to this, SRDenseNet in our architecture can reconstruct HR images of downsampled images of our dataset.
\vspace{-5pt}

\subsection{Peak Signal-to-Noise Ratio}
To compare the quality of the reconstructed image with the original one, we use the Peak Signal-to-Noise Ratio (PSNR) measurement technique. It is given for grey-scale images by the equation:
\begin{linenomath}
\begin{equation}
PSNR = 10 {.} log_{10} \begin{bmatrix} \frac{I^2}{MSE} \end{bmatrix}
\end{equation}
\end{linenomath}
where MSE is the mean square error which is given by the equation below:
\begin{linenomath}
\begin{equation}
MSE = \frac{1}{N {.} 
M} \sum_{i=1}^N \sum_{j=1}^M (y_{ij} - \overline{y}_{ij})
\end{equation}
\end{linenomath}
$I$ is the maximum intensity value of the image, which is 255 in this work. $N$ is the number of rows and $M$ is the number of columns. $y$ is the original image and $\overline{y}$ is the reconstructed~image.
\section{Results and Discussion}\label{444}
In this section, we compare  the numerical results of the applications of INR-based compression techniques with and without our architecture in terms of reconstruction quality, training speed, and GPU adaptivity. In this work, we examine an INR-based method, SIREN with two, three, and four MLP layers. Tables \ref{tab1}--\ref{tab3} show the size of the MLPs with two, three, and four layers, respectively, in detail.
\begin{table}[H]

\caption{The estimation of SIREN network size with two layers. \label{tab1}}
\newcolumntype{C}{>{\centering\arraybackslash}X}
\begin{tabularx}{\textwidth}{CCC}
\toprule
\textbf{Layer (Type)}	& \textbf{Output Shape}	& \textbf{Number of Parameters}\\
\midrule
Linear-1 & [$-$1, 1, 128] & 512\\
Linear-2 & [$-$1, 1, 128] & 16,512\\
Linear-3 & [$-$1, 1, 1] & 129\\
\midrule
Total params: 17,153 &&\\
Trainable params: 17,153 &&\\
Non-trainable params: 0 &&\\
\midrule
Input size (MB): 0.00 &&\\
Forward/backward pass size (MB): 0.00 &&\\
Params size (MB): 0.07 &&\\
Estimated total size (MB): 0.07 &&\\
\bottomrule
\end{tabularx}
\end{table}
\unskip
\begin{table}[H]
\caption{The estimation of SIREN network size with three layers. \label{tab2}}
\newcolumntype{C}{>{\centering\arraybackslash}X}
\begin{tabularx}{\textwidth}{CCC}
\toprule
\textbf{Layer (Type)}	& \textbf{Output Shape}	& \textbf{Number of Parameters}\\
\midrule
Linear-1 & [$-$1, 1, 128] & 512 \\
Linear-2 & [$-$1, 1, 128] & 16,512\\
Linear-3 & [$-$1, 1, 128] & 16,512\\
Linear-4 & [$-$1, 1, 1] & 129\\
\midrule
Total params: 33,665 &&\\
Trainable params: 33,665 &&\\
Non-trainable params: 0 &&\\
\midrule
Input size (MB): 0.00 &&\\
Forward/backward pass size (MB): 0.00 &&\\
Params size (MB): 0.13 &&\\
Estimated total size (MB): 0.13 &&\\
\bottomrule
\end{tabularx}
\end{table}
\unskip
\begin{table}[H]
\caption{The estimation of SIREN network size with four layers.\label{tab3}}
\newcolumntype{C}{>{\centering\arraybackslash}X}
\begin{tabularx}{\textwidth}{CCC}
\toprule
\textbf{Layer (Type)}	& \textbf{Output Shape}	& \textbf{Number of Parameters}\\
\midrule
Linear-1 & [$-$1, 1, 128] & 512 \\
Linear-2 & [$-$1, 1, 128] & 16,512\\
Linear-3 & [$-$1, 1, 128] & 16,512\\
Linear-4  &             [$-$1, 1, 128]   &       16,512\\
Linear-5  &              [$-$1, 1, 1]    &         129\\

\midrule
Total params: 50,177 &&\\
Trainable params: 50,177 &&\\
Non-trainable params: 0 &&\\
\midrule
Input size (MB): 0.00 &&\\
Forward/backward pass size (MB): 0.00 &&\\
Params size (MB): 0.19 &&\\
Estimated total size (MB): 0.20 &&\\
\bottomrule
\end{tabularx}
\end{table}

\subsection{Dataset} In this paper, we are concerned with both 2D and 3D images. We use a volumetric dataset, containing Human CT scan slices obtained from the Visible Human project dataset (\url{https://www.nlm.nih.gov}, {(accessed on 1 April 2022)}
). The database constitutes 463~DICOM axial CT scan slices of a human head with a resolution of 512~{$\times$}~512 pixels and each pixel is made up of 12 bits of grey tone. The thickness of each slice is 0.5 mm.
\subsection{Using SIREN with Our Architecture}
\begin{figure}[H]
\vspace{-9pt}
\includegraphics[width=12 cm]{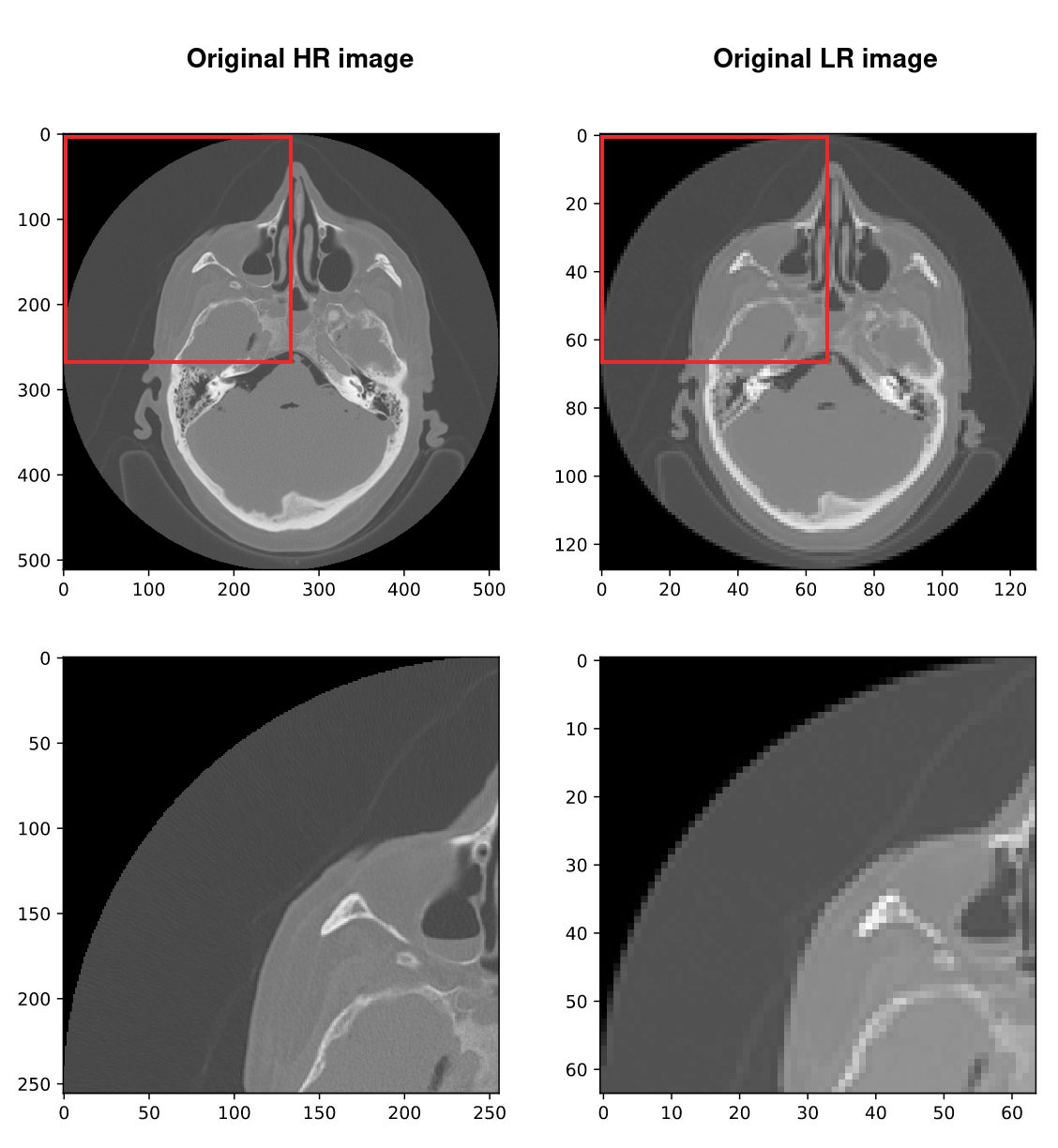}
\caption{Left column shows the high-resolution slices with a size of 512~{$\times$}~512 and the right is the low-resolution slices with a size of 128~{$\times$}~128. Low-resolution slices were obtained by applying
Lanczos resampling on the high-resolution slices.\label{fig3}}
\end{figure}

Based on our architecture, we use the LANCZOS resampling technique to downsample our high-resolution volumetric medical images (512~{$\times$}~512~{$\times$}~463) to (128~{$\times$}~128~{$\times$}~115). This technique decreases the number of voxels from 121,372,672 to just 1,884,160. For simplicity, we show the results on a selected 2D slice, but it should be considered that it can be expanded to 3D data. Figure \ref{fig3} shows a selected slice that is downsampled from 512~{$\times$}~512~{$\times$}~1 to 128~{$\times$}~128~{$\times$}~1 using LANCZOS, which decreases the number of voxels from 262,144 to just 16,384.
\vspace{-5pt}

\begin{figure}[H]
\

\includegraphics[width=13.6 cm]{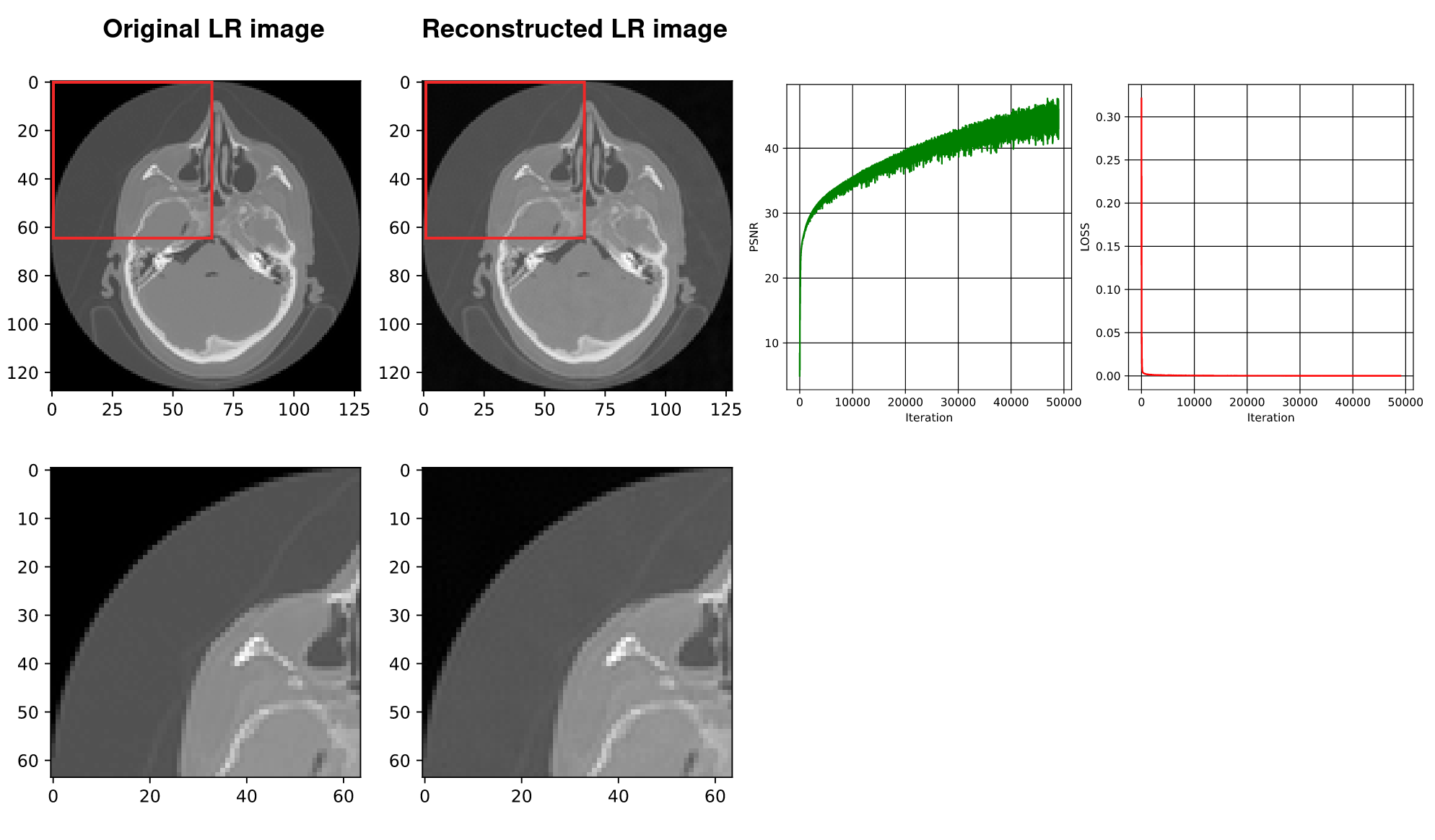}
\caption{{From} 
left, the first image shows the original down-sampled image. The second image illustrates the reconstructed counterpart of the down-sampled image using SIREN.
The next image shows the plot of the PSNR while training the network (best PSNR: 45.38)
and the last image shows the plot of the loss values during the SIREN training (loss:
$1.4805471 \times 10^{-5}$). 
\label{2layersiren}}
\end{figure}

\begin{figure}[H]
\includegraphics[width=13.6 cm]{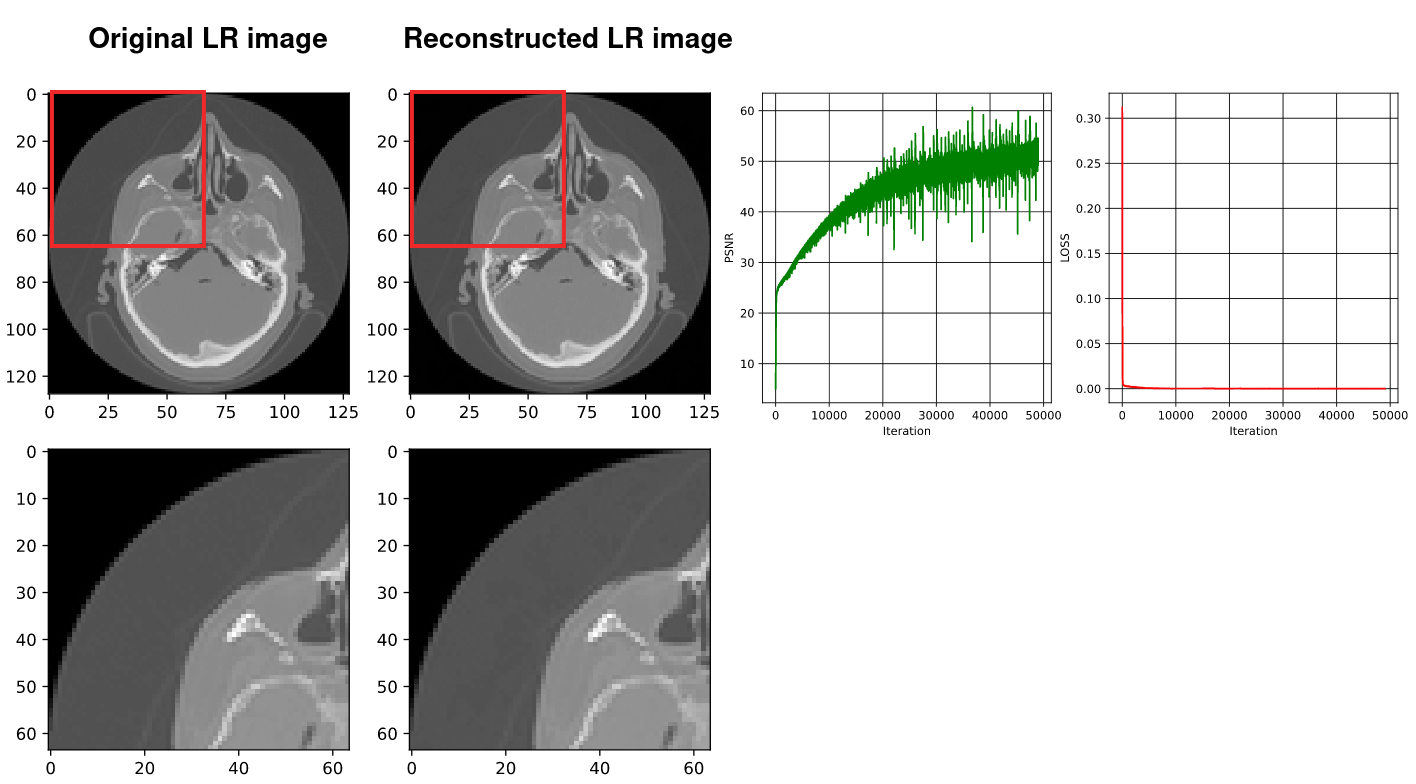}
\caption{{From} 
the left, the first image shows the original down-sampled image. The second image illustrates the reconstructed counterpart of the down-sampled image using SIREN.
The next image shows the plot of the PSNR while training the network (best PSNR: 48.29)
and the last image shows the plot of the loss values during the SIREN training (loss:
1.0805471 $\times \,10^{-5}$).\label{3layersiren}}
\end{figure}

\begin{figure}[H]
\

\includegraphics[width=13.65 cm]{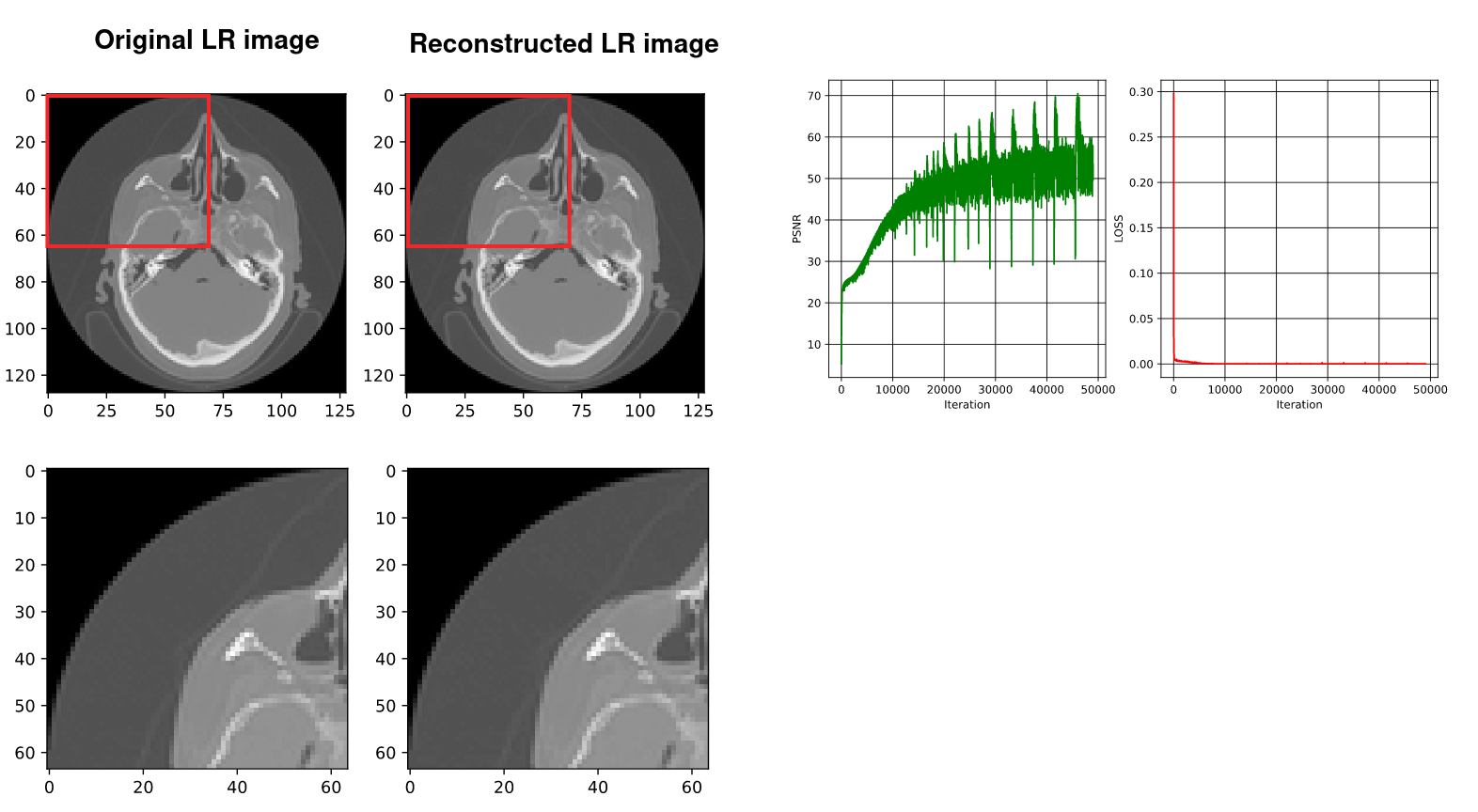}
\caption{{From} 
\textls[+25]{the left, the first image shows the original down-sampled image. The second image illustrates the reconstructed counterpart of the down-sampled image using a four-layer SIREN.
The next image shows the plot of the PSNR during training of the network (best PSNR: 70.63)
and the last image shows the plot of the loss values during the four-layer SIREN training (loss:  8.6481705 $\times \, 10^{-8}$).}\label{4layersiren}}
\end{figure}

\begin{figure}[H]
\includegraphics[width=13.6 cm]{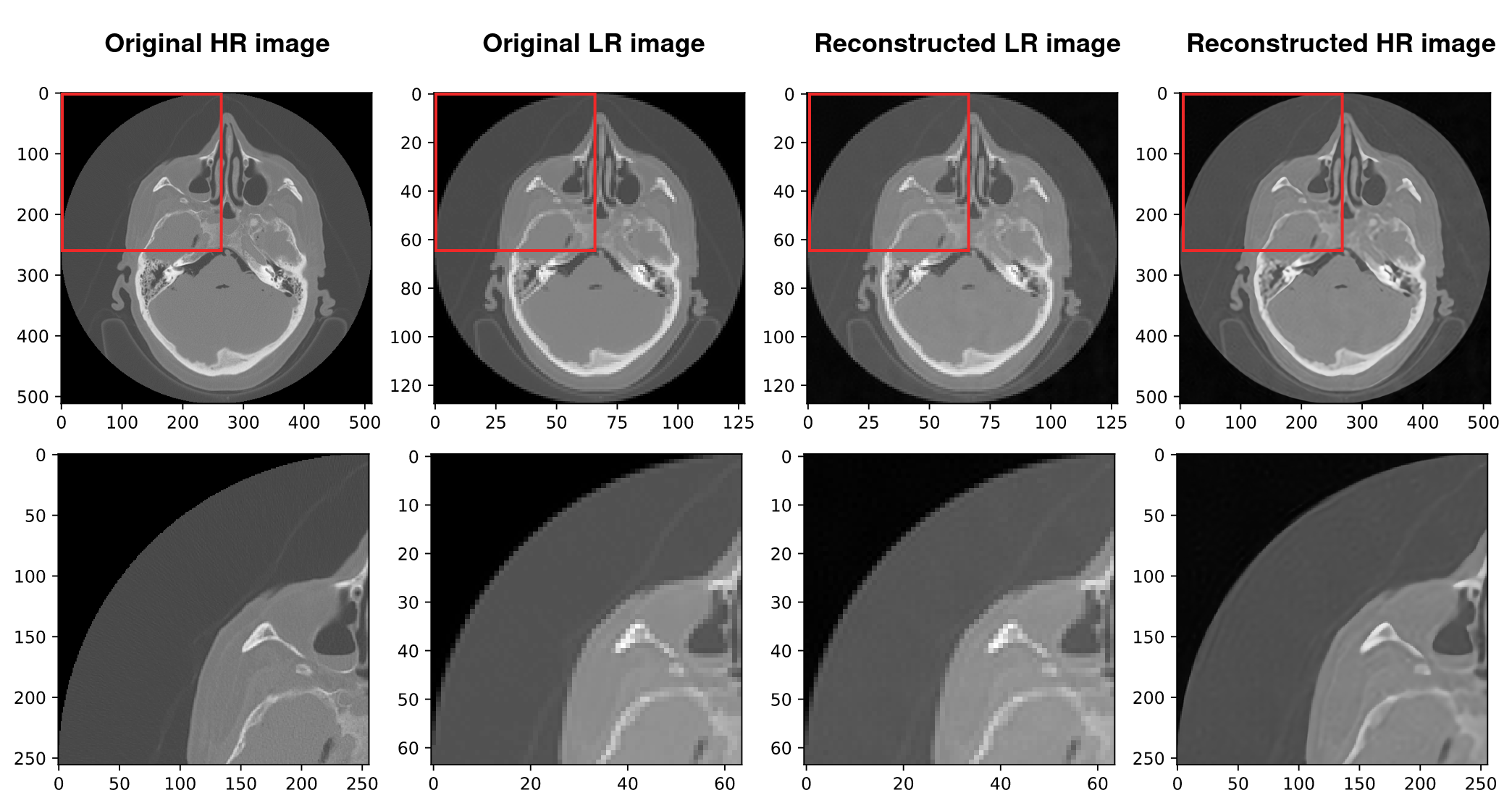}
\caption{\textls[+25]{Results of the whole procedure of our architecture. From the left, the first image
shows the original high-resolution (HR) image. The second image shows the original low resolution (LR). The third image shows the reconstructed low resolution (LR) and the last shows the
reconstructed high resolution (HR). The best PSNR of reconstructed HR in comparison
with the original HR is 34.670.}\label{2layerarch}}
\end{figure}

\begin{figure}[H]
\

\includegraphics[width=13.6 cm]{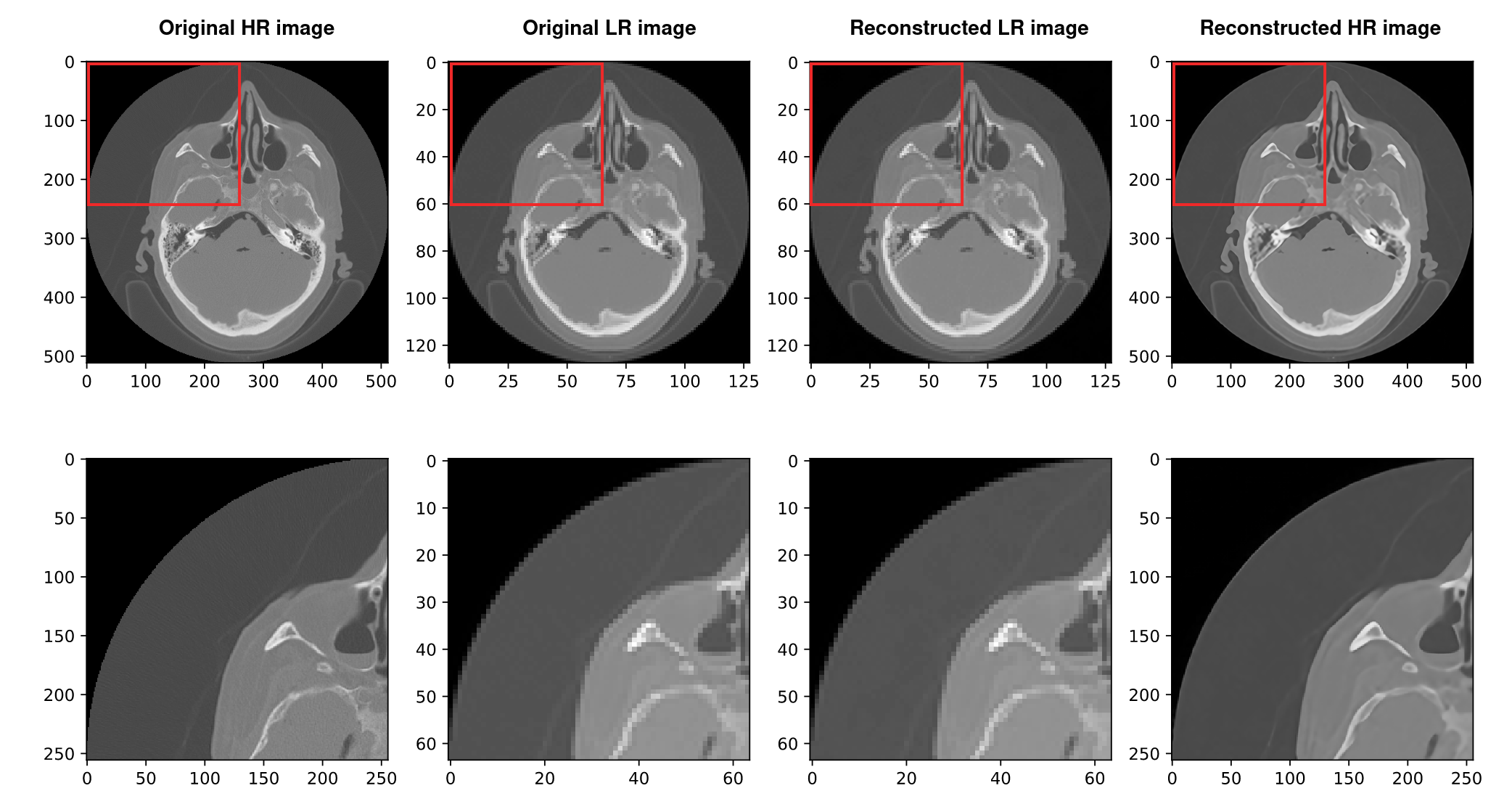}
\caption{\textls[+15]{Results of the whole procedure of our architecture. From the left, the first image
shows the original high-resolution (HR) image. The second shows the original low resolution
(LR). The third image shows the reconstructed low resolution (LR) and the last shows the
reconstructed high resolution (HR). The best PSNR of reconstructed HR in comparison
with original HR is 34.865}.\label{3layerarch}}
\end{figure}

\begin{figure}[H]
\includegraphics[width=13.5 cm]{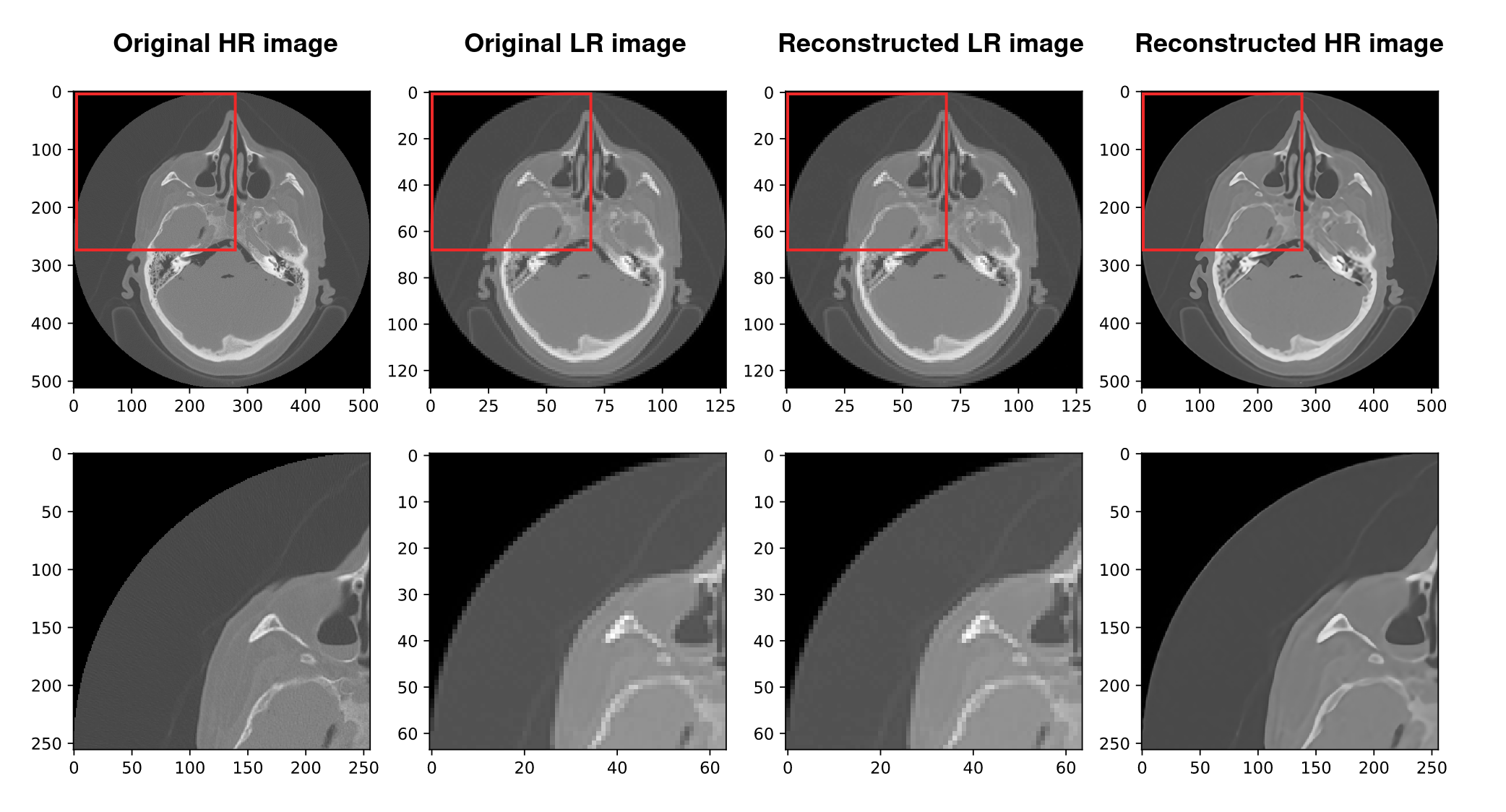}
\caption{\textls[+15]{Results of the whole procedure of our architecture. From the left, the first image
shows the original high-resolution (HR) image. The second shows the original low resolution
(LR). The third image shows the reconstructed low resolution (LR) and the last shows the
reconstructed high resolution (HR). The best PSNR of reconstructed HR in comparison
with the original HR is 35.140}.\label{4layerarch}}
\end{figure}

After the downsampling stage, our architecture uses SIREN with 2, 3, and 4 layers and 128 neurons for each layer. {Figures}~\ref{2layersiren}--\ref{4layersiren} show the results, respectively. The Adam was selected as the optimizer with a learning rate of 0.0015 and batch size of $2^{16}$. MSE was used as the loss function of the MLP. In order to reconstruct the HR slice, we used SRDenseNet and applied it to reconstructed LR images. We also trained it on this kind of CT data in advance so that the trained SRDenseNe is suitable for the targeted class. The final results of using SIREN with our architecture are shown in {Figures}~\ref{2layerarch}--\ref{4layerarch}, respectively. Table~\ref{tab4} compares the results of the SIREN application with 2, 3 and 4 layers, using our architecture.
\vspace{-15pt}

\vspace{-10pt}

\begin{table}[H]
\caption{Shows comparison results of the implementation of SIREN with our architecture in terms of quality, speed, GPU memory allocation, and compression rate with 2, 3, and 4 layers of SIREN.\label{tab4}}
\newcolumntype{C}{>{\centering\arraybackslash}X}
\begin{tabularx}{\textwidth}{CCcCC}
\toprule
\textbf{Number of Layers}	& \textbf{Best PSNR}	& \textbf{Training Time(s)/50,000 Iters} & \textbf{Compression Rate} & \textbf{GPU Memory (KB)}\\
\midrule
Two Layer & 34.670 & 55.76 & 3.65 & 1038\\
Three Layer & 34.865 & 74.80 & 1.96 & 1296\\
Four Layer & 35.140 & 99.61 & 1.28 & 1554\\
\bottomrule
\end{tabularx}
\end{table}

\vspace{-15pt}

\vspace{-9pt}

\subsection{Using SIREN without Our Architecture~\cite{COM1}}
\begin{figure}[H]
\

\includegraphics[width=13.5 cm]{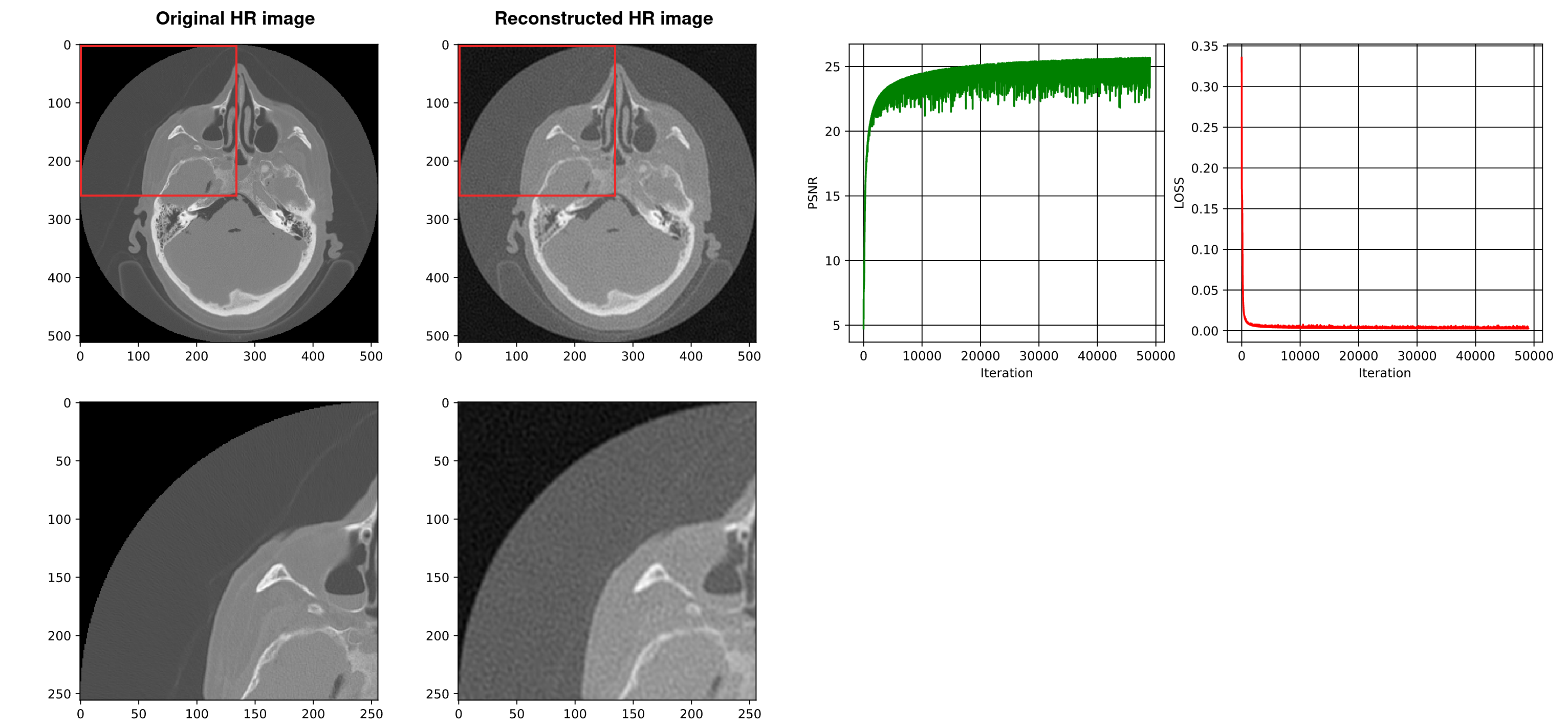}
\caption{{Shows} the reconstruction of HR images using a two-layer SIREN without our architecture. The best PSNR is 25.26.\label{SIREN-2}}
\end{figure}

\begin{figure}[H]
\includegraphics[width=13.5 cm]{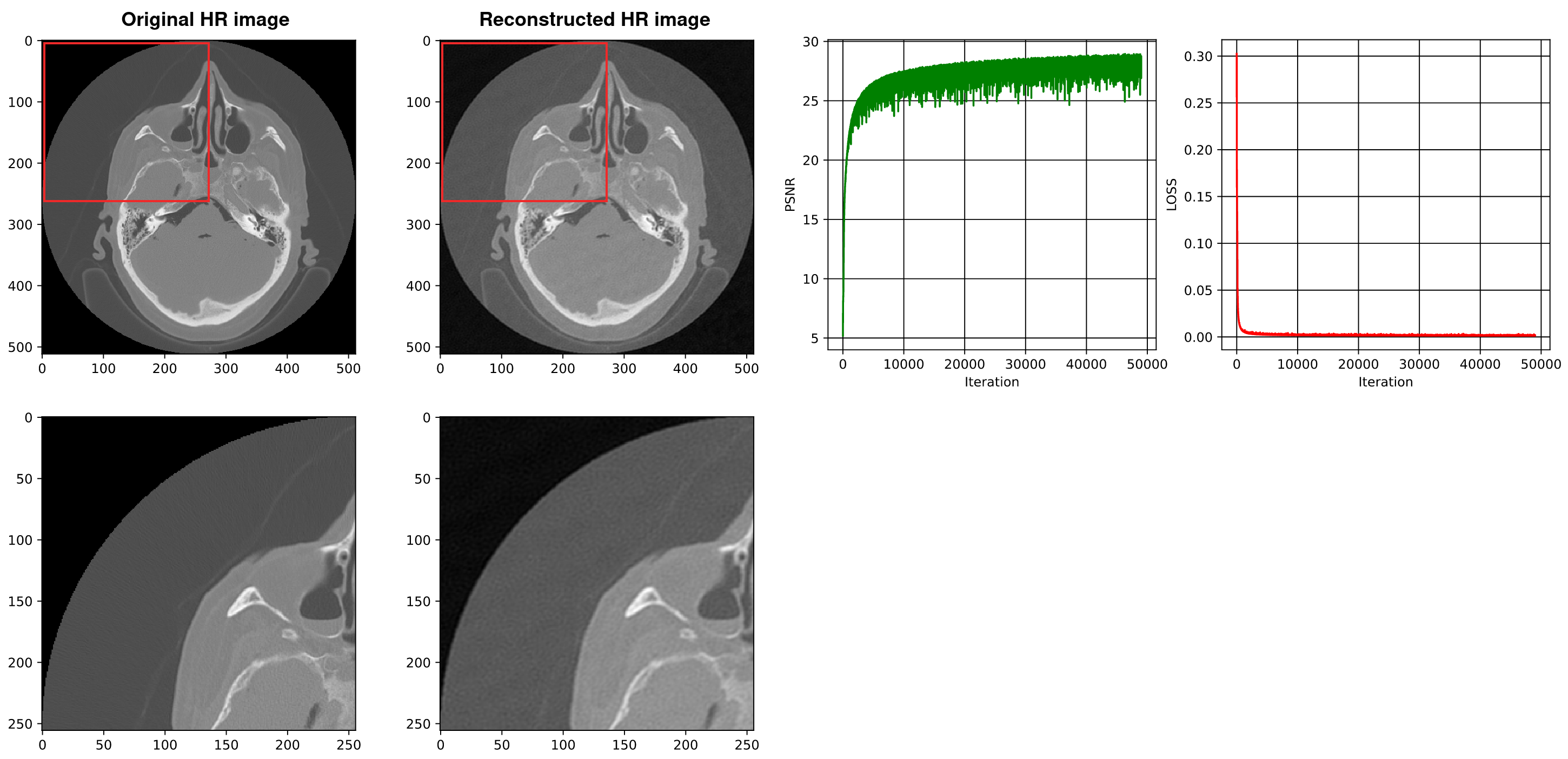}
\caption{{Shows} the reconstruction of HR images using a three-layer SIREN without our architecture. The best PSNR is 28.80.\label{SIREN-3}}
\end{figure}

\begin{figure}[H]
\includegraphics[width=13.5 cm]{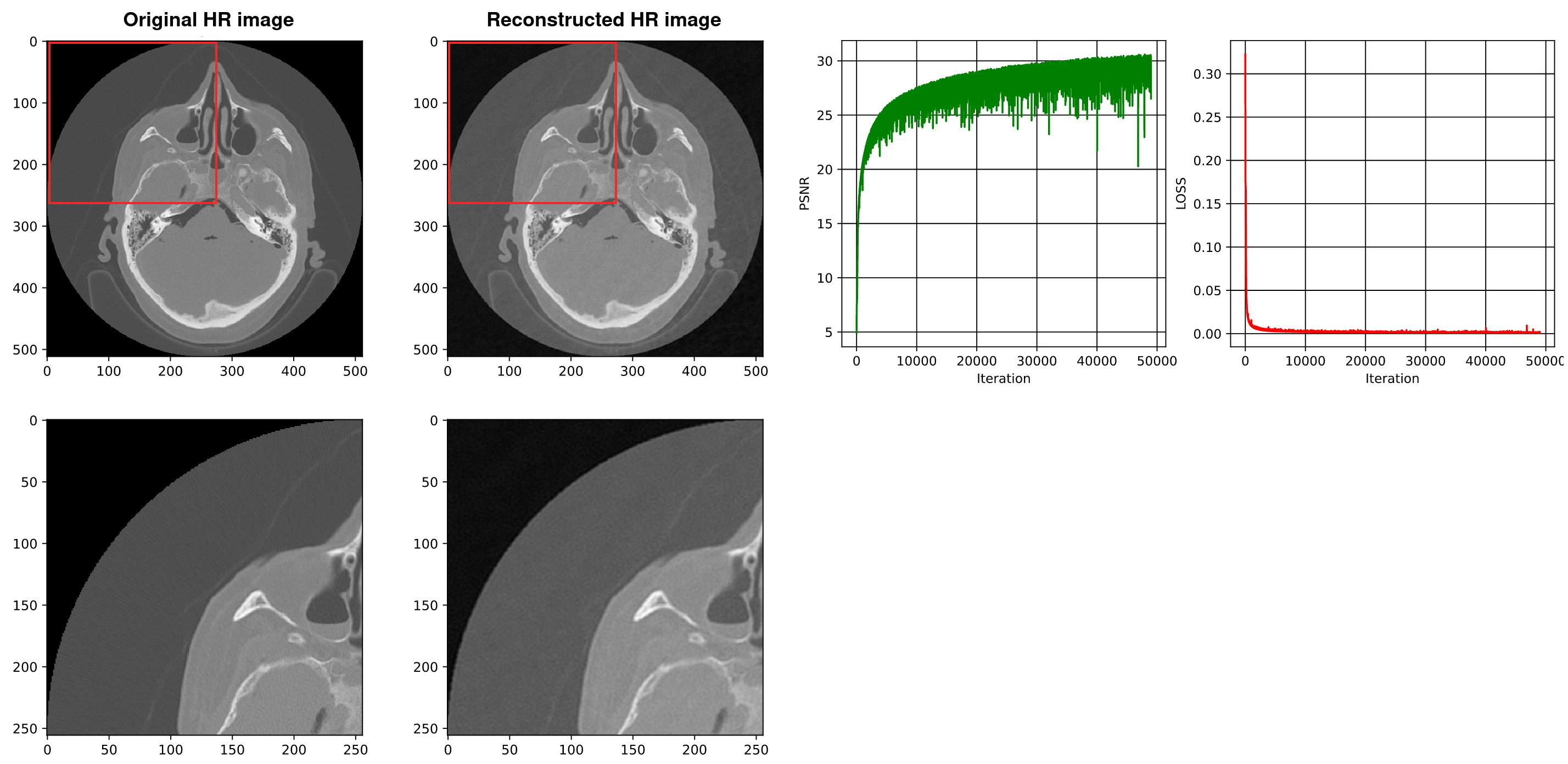}
\caption{{Shows} the reconstruction of HR images using a four-layer SIREN without our architecture. The best PSNR is 30.68.\label{SIREN-4}}
\end{figure}

In this section, we show the results of SIREN implementation without our architecture which was implemented in the Ref.~\cite{COM1}. Figures~\ref{SIREN-2}--\ref{SIREN-4} show the results of high-resolution (HR) reconstruction of 2, 3 and 4 layers of SIREN without our architecture, respectively. Table~\ref{tab5} illustrates the results of using SIREN without our architecture with 2, 3, and 4~layers.

\vspace{-16pt}

\vspace{-10pt}

\begin{table}[H]
\caption{Shows comparison results of the implementation of SIREN without our architecture in terms of quality, speed, GPU memory allocation, and compression rate with 2, 3, and 4 layers with SIREN.\label{tab5}}
\newcolumntype{C}{>{\centering\arraybackslash}X}
\begin{tabularx}{\textwidth}{CcCCC}
\toprule
\textbf{Number of Layers}	& \textbf{Best PSNR}	& \textbf{Training Time(s)/\linebreak50,000 Iters} & \textbf{Compression Rate} & \textbf{GPU Memory (KB)}\\
\midrule
Two Layer & 25.269 & 664.92 & 3.65 & 10,254\\
Three Layer & 28.800 & 1004.21 & 1.96 & 10,512\\
Four Layer & 30.689 & 1336.53 & 1.28 & 10,770\\
\bottomrule
\end{tabularx}
\end{table}

\vspace{-10pt}

\unskip

\unskip

\unskip

\subsection{Comparison with Existing Methods}
\begin{figure}[H]
\

\includegraphics[width=10.5 cm]{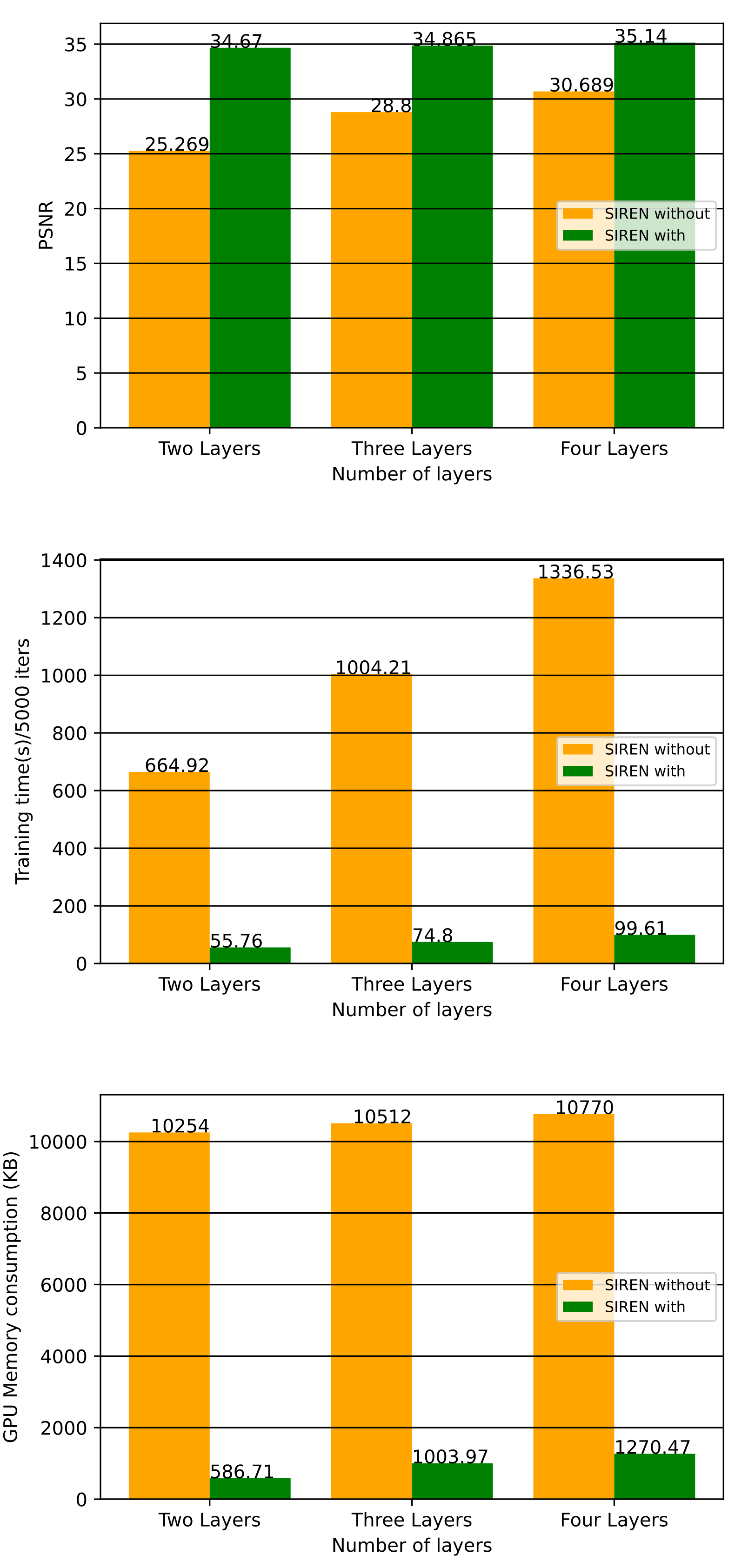}
\caption{{Bar} 
charts compare the results of using SIREN without our architecture~\cite{COM1} and with our architecture. As can be seen, SIREN implementation with our architecture outperforms using SIREN without our architecture~\cite{COM1} in terms of quality, speed, and GPU memory allocation with 2, 3, and 4 MLP layers. \label{Plots}}
\end{figure}

\begin{figure}[H]
\begin{adjustwidth}{-\extralength}{0cm}
\

\centering
\includegraphics[width=13.5cm]{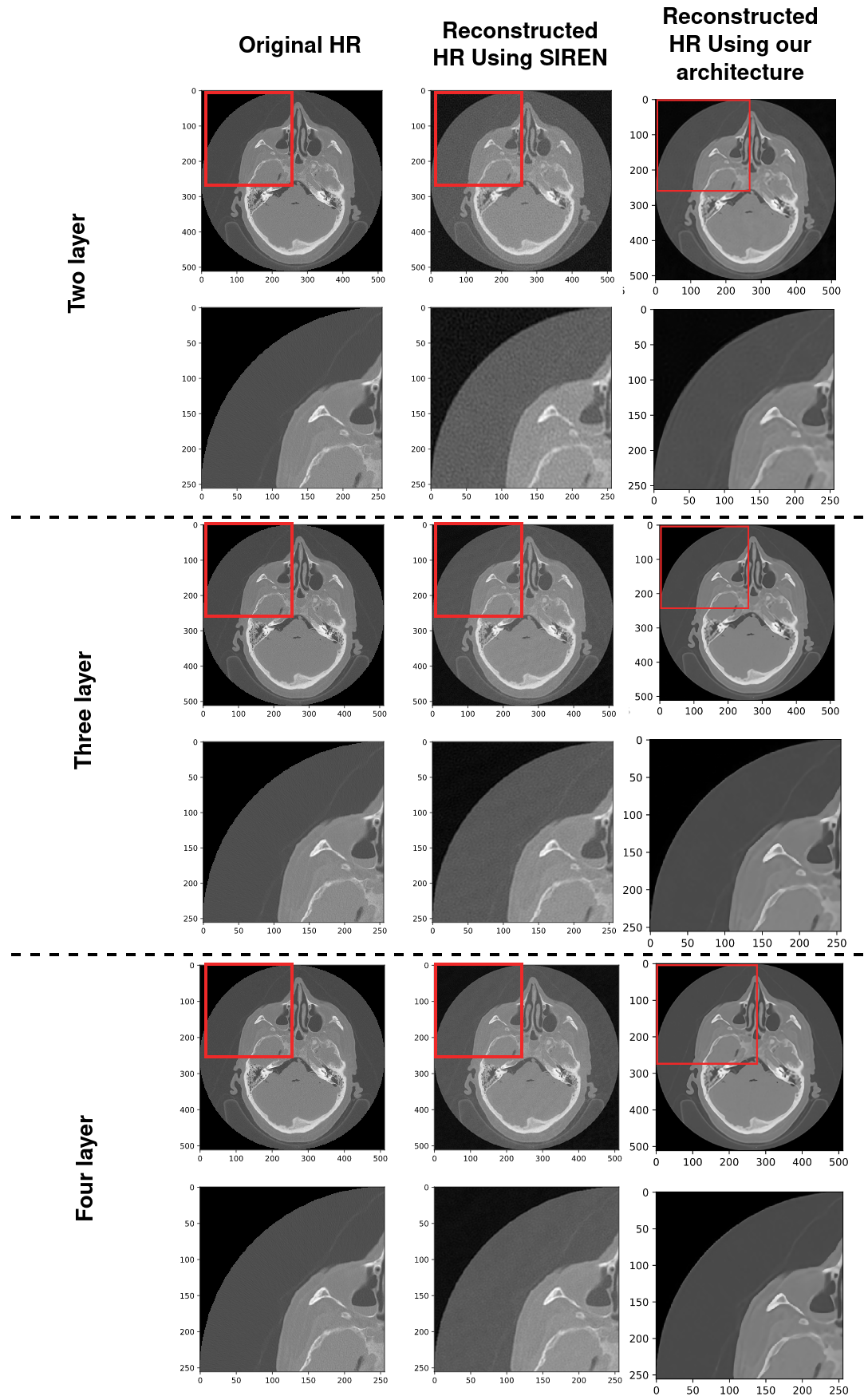}
\end{adjustwidth}
\caption{Comparison of the quality of reconstructed HR using SIREN in the Ref.~\cite{COM1} and using SIREN with our architecture with 2, 3 and 4 MLP layers.\label{Total}}
\end{figure}

In this section, we compare the performance of SIREN with our architecture and SIREN without our architecture which is used in the Ref.~\cite{COM1} with 2, 3 and 4 layers and 128 neurons for each layer. From the general perspective, both methods use SIREN to compress volumetric medical images. In the Ref.~\cite{COM1} they applied SIREN without any downsampling and upsampling techniques which led to high GPU memory consumption and low training speed.  In Figure~\ref{Plots} there are three bar charts, showing that SIREN with our architecture outperforms SIREN without our architecture in the Ref.~\cite{COM1} in terms of quality, speed, and GPU memory consumption with 2, 3, and 4 MLP layers. As can be seen, there are considerable gaps between our technique and SIREN in the Ref.~\cite{COM1} in terms of training speed and GPU memory consumption. The PSNR gap is decreased by increasing the number of layers, but our architecture still outperforms the SIREN in the Ref.~\cite{COM1}. Table~\ref{tab6} compares the results of using SIREN to compress volumetric medical images in the Ref.~\cite{COM1} with using SIREN with our architecture in three factors of PSNR, training speed and GPU memory consumption. Figure~\ref{Total} compares the final results of using SIREN in the Ref.~\cite{COM1} and using SIREN with our architecture with 2, 3 and 4 layers. Table \ref{tab8} shows the comparison of the results of Shen's~\cite{shen}, Mishra's~\cite{Autoencoder}, SIREN~\cite{COM1} and ours in terms of PSNR.

\begin{table}[H]
\caption{{Display and comparison of the final results that were obtained using SIREN in the Ref.~\cite{COM1} and using SIREN with our architecture. As it can be seen, using SIREN with our architecture outperforms using SIREN in the Ref.~\cite{COM1} in three terms of quality, speed, and GPU memory allocation with 2, 3 and 4 MLP layers}.\label{tab6}}
\begin{adjustwidth}{-\extralength}{0cm}
\newcolumntype{C}{>{\centering\arraybackslash}X}
\begin{tabularx}{\fulllength}{ccCCC}
\toprule
\textbf{} & \textbf{Number of Layers}	& \textbf{SIREN without Our Architecture~\cite{COM1}}	& \textbf{SIREN with Our Architecture}\\
\midrule

\multirow[m]{3}{*}{Best PSNR (dB)}    & 2 layers			& 25.269			& \textbf{{34.670}}\\
& 3 layers			& 28.800			& \textbf{34.865}\\
& 4 layers			& 30.689			& \textbf{35.140}\\
\midrule
\multirow[m]{3}{*}{Training time(s) (for 15000 iters)}    & 2 layers			& 664.92			& \textbf{55.76}\\
& 3 layers			& 1004.21			& \textbf{74.80}\\
& 4 layers			& 1336.53			& \textbf{99.61}\\
\midrule
\multirow[m]{3}{*}{GPU memory consumption (KB)}    & 2 layers			& 10,254			& \textbf{1,038}\\
& 3 layers			& 10,512			& \textbf{1,296}\\
& 4 layers			& 10,770			& \textbf{1,554}\\
\bottomrule
\end{tabularx}
\end{adjustwidth}
\end{table}  

\vspace{-12pt}

\begin{table}[H]
\caption{Comparison of results of our technique with SIREN~\cite{COM1}, Mishra's~\cite{Autoencoder} and Shen's~\cite{shen} in terms of PSNR.\label{tab8}}
\newcolumntype{C}{>{\centering\arraybackslash}X}
\begin{tabularx}{\textwidth}{CC}
\toprule
\textbf{Methods}	& \textbf{Best PSNR}\\
\midrule
Shen's~\cite{shen} & 27.50 \\
Mishra's~\cite{Autoencoder} & 29.33\\

SIREN~\cite{COM1} (2 layers)  & 25.269 \\
SIREN~\cite{COM1} (3 layers)  & 28.800 \\
SIREN~\cite{COM1} (4 layers)  & 30.689 \\
Ours (2 layers)  & \textbf{34.670}\\
Ours (3 layers)  & \textbf{34.865}\\
Ours (4 layers)  & \textbf{35.140}\\
\bottomrule
\end{tabularx}
\end{table}
\unskip

\section{Conclusions}\label{555}

Our architecture significantly outperforms other INR-based techniques without our architecture, which are regular in medical image compression. The experiments show that
our proposed architecture is a novel implicit neural representation of medical volume data, reaching
both a high compression rate and high quality of reconstruction. The architecture is
simple and re-configurable. It may replace volume data and be regarded as a new representation form. Moreover, the three modules can be replaced by others according to different
applications. For example, manual annotation may work on low-resolution volume to save
time, while rendering may work on high-resolution volume for accuracy.

The quality of our reconstructed high-resolution images with a small version of SIREN is
considerably higher than direct SIREN with the same size. We also note that the compression rate depends on the deep network structure. To reach a high compression rate and
low loss for volume data, the SR module is crucial. Moreover, the generalization of the SR module
is another concern. These are worth our effort in the future.

\vspace{6pt}

\authorcontributions{{Conceptualization, A.S. and H.Y.; methodology, A.S.; software, A.S.; validation, A.S. and H.Y.; formal analysis, A.S.; investigation, A.S. and H.Y.; resources, A.S. and H.Y.; data curation, A.S. and H.Y.; writing---original draft preparation, A.S.; writing---review and editing, A.S. and H.Y.; visualization, A.S.; supervision, H.Y.; project administration, H.Y.; funding acquisition, H.Y. All authors have read and agreed to the published version of the manuscript.}}
\funding{This research received no external funding.}
\institutionalreview{Not applicable.}

\informedconsent{{Not applicable.}}

\dataavailability{\url{https://www.nlm.nih.gov/research/visible/visible_human.html}, {(accessed on 1 April 2022).} 
}

\conflictsofinterest{The authors declare no conflict of interest.}

\abbreviations{Abbreviations}{
The following abbreviations are used in this manuscript:\\

\noindent
\begin{tabular}{@{}ll}
INR & Implicit Neural Representation\\
MLP & Multi-Layer Perceptron\\
CT & Computed Tomography\\
SIREN &  Sinusoidal representation network\\
\end{tabular}
}

\begin{adjustwidth}{-\extralength}{0cm}

\reftitle{References}

\PublishersNote{}
\end{adjustwidth}
\end{document}